\documentclass{PoS}

\usepackage{amsmath}
\usepackage{bm}
\usepackage[vcentermath]{youngtab}
\title{How Many Degrees of Freedom Has the Gluon ?}

\ShortTitle{How Many Degrees of Freedom Has the Gluon ?}

\author{\speaker{Vincent Mathieu}\thanks{IISN Scientific Research Worker}\\
        Service de Physique Nucl\'{e}aire et Subnucl\'{e}aire, Universit\'{e} de Mons, Acad\'{e}mie universitaire Wallonie-Bruxelles, Place du Parc 20, B-7000 Mons, Belgium\\
        E-mail: \email{vincent.mathieu@umons.ac.be}}

\abstract{In constituent models, glueballs are described as bound states of effective gluons. The dynamical mass generation lead to a gluon mass without violating gauge invariance. One can then consider three degrees of freedom for those effective massive gluons. However, models with only two degrees of freedom are shown to be in better agreement with the lattice glueball spectrum. I review both models for two- and three-gluon models and explain why even though the gluon gains a mass, it behaves as a massless particle with only two degrees of freedom.}

\FullConference{International Workshop on QCD Green's Functions, Confinement, and Phenomenology - QCD-TNT09\\
		 September 07 - 11 2009\\
		 ECT Trento, Italy}

\begin{document}

\section{Introduction}
The gluon is the fundamental particle of the Yang-Mills theory. Gauge invariance forbids a mass term in Lagrangian. The gluon is therefore a massless particle. However it has been argued long time ago~\cite{Cornwall:1981zr} that non-perturbative effects might lead to a dynamical mass for the gluon without breaking gauge invariance.

Massive and massless particles are different representations of the Poincar\'{e} group. Massless particles can not be at rest and only the two maximally spin projections are allowed. {\it A contrario}, massive spin-$s$ particles have $2s+1$ polarisation states. The gluon, a vector particle, has intrinsically two degrees of freedom but, in view of the its non-trivial dispersion relation, one can wonder if the gluon gains a third polarization.

Those group theory considerations are only valid for free particles. The problem of the number of gluonic degrees of freedom could seem ill-defined since gluons are never free particles.
However, beside gluon mass generation, gluon self-couplings induce the existence of bound states, called glueballs. Glueballs are well described by bound states of two, three or more constituent gluons~\cite{Mathieu:2008me}. The problem could be then understood as with how many degrees of freedom should we model effective gluons to have a coherent description of glueballs.

Those effective gluons may have two of three polarisation states if we think about massless or massive particles. As we will see, the glueball spectrum depends of the number of gluonic degrees of freedom. It could be then possible to answer to our question by resorting to glueball spectroscopy. I will then explore the differences of models for glueballs with two and three polarisation states for the constituent gluons. Due to the lack of experimental candidates for glueballs, I will compare both approaches with quarksless Quantum Chromodynamics (QCD) spectrum obtained on a lattice.

In section 2, I review the concept of the gluon mass generation. The simplest way to incorporate this feature in a glueball formalism is to allow three polarizations for the effective gluons. We developed those models for two- and three-gluon glueballs in section~3. We then conclude that this description is not appropriate to compare with the glueball spectrum obtained in lattice QCD. I then review the construction of two-gluon glueballs with only two degrees of freedom for the gluons in section 4. I present also arguments favor this interpretation of lattice results for two-, three and four-gluon glueballs and gluelump in section 5. Section 6 summarizes the conclusions.

\section{Gluon Mass}

The gauge bosons are massless at the Lagrangian level but there are hints that
they obey massive dispersion relations. Gauge invariance through the Ward identity $k^\mu\Pi_{\mu\nu}(k)=0$ forbids the apparence of a mass term in all order in perturbation theory. However, a pole in the the self-energy would imply a mass term in the propagator without breaking gauge invariance~\cite{Jackiw:1973tr}. This is the Schwinger mechanism~\cite{Schwinger:1962tp}.

The so called dynamical mass, is defined by the position of the pole
of the dressed gluon propagator. Cornwall arrived to such a
dynamical mass by analyzing the gluon Dyson-Schwinger equations in
the early 80's~\cite{Cornwall:1981zr}. This infinite set of couple
equations cannot be solved analytically. One must resort to a
truncation scheme. By a clever resummation of Feynman diagrams, the pinch technique,
Cornwall found a gauge-invariant procedure to truncate these
equations without spoiling gauge invariance. With this technique, a full gluon propagator in quarkless
QCD emerges
\begin{equation}\label{eq:gluon_prop_cornwall}
d^{-1}(q^2) = \left(q^2+m^2(q^2)\right) bg^2 \ln\left[\frac{q^2+4m^2(q^2)}{\Lambda^2}\right],
\end{equation}
with a dynamical mass
\begin{equation}\label{eq:dym_mass}
    m^2(q^2) =
    m^2\left(\frac{\ln\left[(q^2+4m^2)/\Lambda^2\right]}
    {\ln\left(4m^2/\Lambda^2\right)}\right)^{-12/11}.
\end{equation}
In Eq.~\eqref{eq:gluon_prop_cornwall},  $b=11N/48\pi^2$ is the first
coefficient of the beta function for quarkless QCD. The mass term
that appears has finite value at zero momentum. The gluon mass can
be related to the gluon condensate $\langle0| F_{\mu\nu}^a
F^{\mu\nu}_a|0\rangle$ from which the value $m=(500\pm200)$~MeV
arises. The perturbative pinch technique invented by Cornwall was recently improved and applied for the gluon Dyson-Schwinger equations as a convient truncation scheme~\cite{Binosi:2009qm}. Various solutions for the gluon mass were found where the solution~\eqref{eq:dym_mass} emerged as a particular case~\cite{Aguilar:2007ie}.

Bernard proposed a different definition for the gluon
mass~\cite{Bernard:1981pg}. Consider the potential energy of a pair
of heavy, static sources in the adjoint representation of the color
group. As the separation of the adjoint sources (static gluons) is
increased, the potential will increase linearly as a string or a
flux tube is  formed between them. The energy stored in the string
will at some point be large enough to pop up a pair of dynamical
gluons out of the vacuum. The effective gluon mass is defined as
half of the energy stored in the flux tube at this point.
Monte-Carlo simulations of this phenomenon show a effective gluon
mass in the range 500-800~MeV.

The effective gluon mass was also investigated in the bag model~\cite{Donoghue:1983fy}. Even
though the gluon is massless in the bag model, a net energy of $740 \pm 100$~MeV is required
to produce a gluon due to confinement.

All these arguments support the use of an effective gluon mass to
describe the dynamics of QCD. It is therefore possible to envisage
an approach to bound states made of constituent massive
gluons. Massive representations have more degrees of freedom that massless ones. Gluon with three polarisations should them be used in glueballs. However, the longitudinal component could be eaten somehow by the scalar massless pole that appears to trigger the Schwinger mechanism~\cite{Jackiw:1973tr,Binosi:2009qm}. In the next sections, I review both approaches.

\section{Three Degrees of Freedom}
\subsection{Two-gluon Glueballs}
One of the pioneering works on two-gluon glueballs was the study by Cornwall and
Soni~\cite{Cornwall:1982zn}. The large value of the effective gluon mass led them to propose a nonrelativistic approach to gluonium. They used a confining potential which saturates at
large distances constrained by Bernard's results~\cite{Bernard:1981pg}
\begin{equation}\label{eq:pot_sat}
    V_{C}(r) = 2m\left(1-e^{-r/r_s}\right).
\end{equation}

The construction of the basis when the gluons are spin$-1$ particles is an easy task. The quantum numbers allowed for a system of two gluons with three degrees of freedom are simply given by
\begin{equation}\label{}
    \bm J = \bm L + \bm S, \qquad P = (-1)^L,\qquad C = +,\qquad \text{and }S = \{0,1,2\}.
\end{equation}
Cornwall and Soni  presented results for states with quantum numbers $L=0$, $J^{PC} = 0^{++}, 2^{++}$, and $L=1$, $J^{PC} = 1^{-+}, 2^{-+}$. The presence of vector states is a characteristic of models with three degrees of freedom for the gluons. Those states are forbidden when only two polarization are taken into account.

The OGE potential~\eqref{eq:pot_oge_cornwall} is necessary to have a detailed spectrum. Indeed, without OGE, the confining potential~\eqref{eq:pot_sat} gives rise to degenerate scalar and tensor ($L=0$) and $L=1$ states. The Coulomb and spin-dependent interactions at short-range were derived from a nonrelativistic expansion of the Feynman graphs
for two-gluon scattering. They considered massive exchanged gluons
(with the same mass as the constituent one) to keep the gauge
invariance of the amplitudes to the given order. This one-gluon
exchange (OGE) potential involves Yukawa, spin-orbit, spin-spin and
tensor forces,
\begin{equation}\label{eq:pot_oge_cornwall}\begin{split}
V_{\text{oge}}(\bm r) =& -\frac{\lambda e^{-mr}}{r}\left(\frac{2s-7m^2}{6m^2}+\frac{1}{3}\bm
S^2\right) + \frac{\pi\lambda\delta(\bm r)}{3m^2} \left(\frac{4m^2-2s}{m^2}+\frac{5}{2}\bm
S^2\right) \\
&- \frac{3\lambda}{2m^2}\bm L\cdot\bm
S\frac{1}{r}\frac{\partial}{\partial r} \frac{e^{-mr}}{r} +
\frac{\lambda}{2m^2}\left[\left(\bm S\cdot\bm\nabla\right)^2
-\frac{1}{3}\bm S^2\bm\nabla^2\right]\frac{e^{-mr}}{r}.
\end{split}
\end{equation}
$s$ is the glueball mass squared, which we can set to $s=4m^2$ in a first approximation, and
$\lambda=3\alpha_s$ is the strong coupling constant.

\begin{figure}[htb]\centerline{
\includegraphics[width=0.45\linewidth]{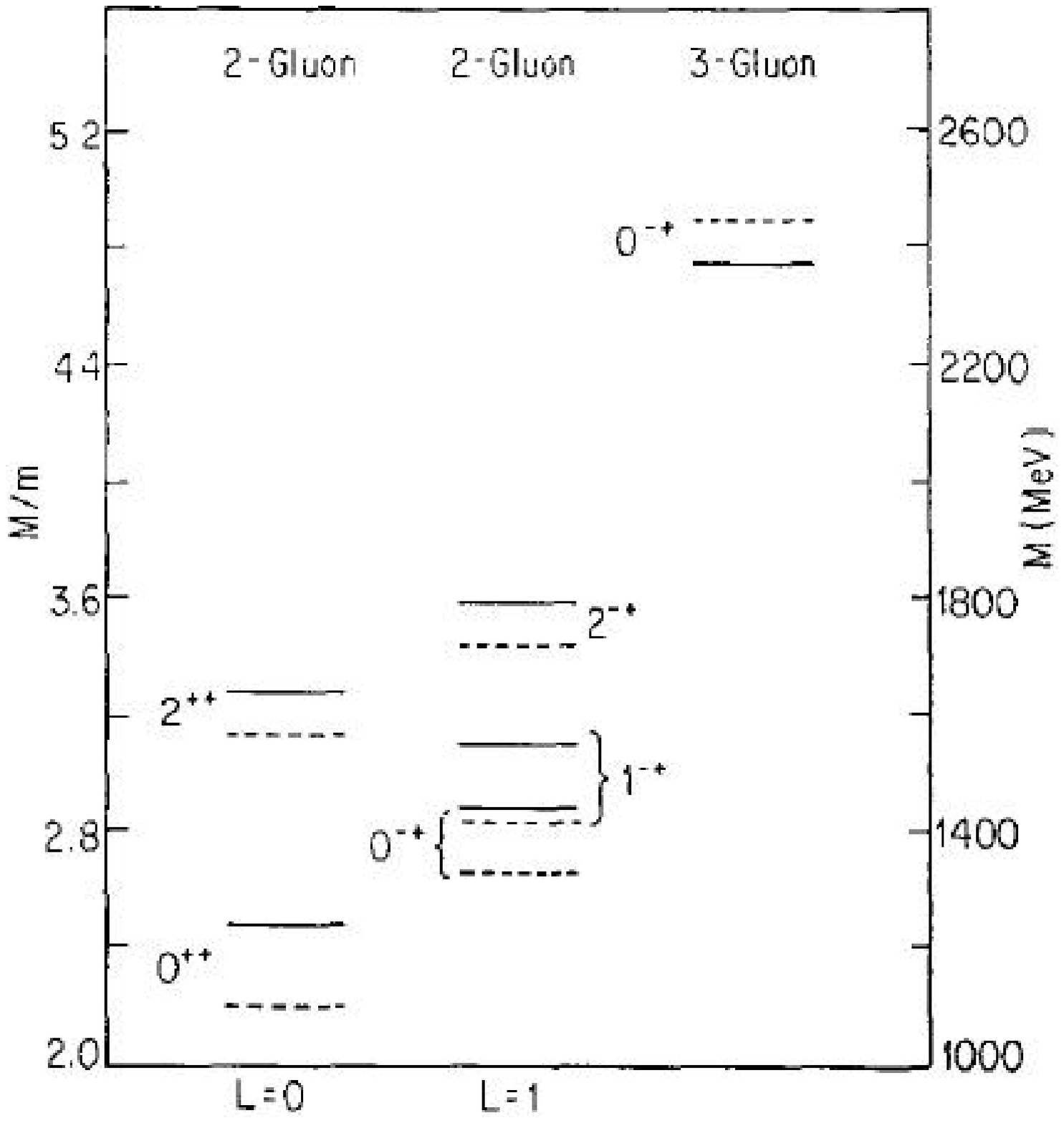}
\includegraphics[width=0.5\linewidth]{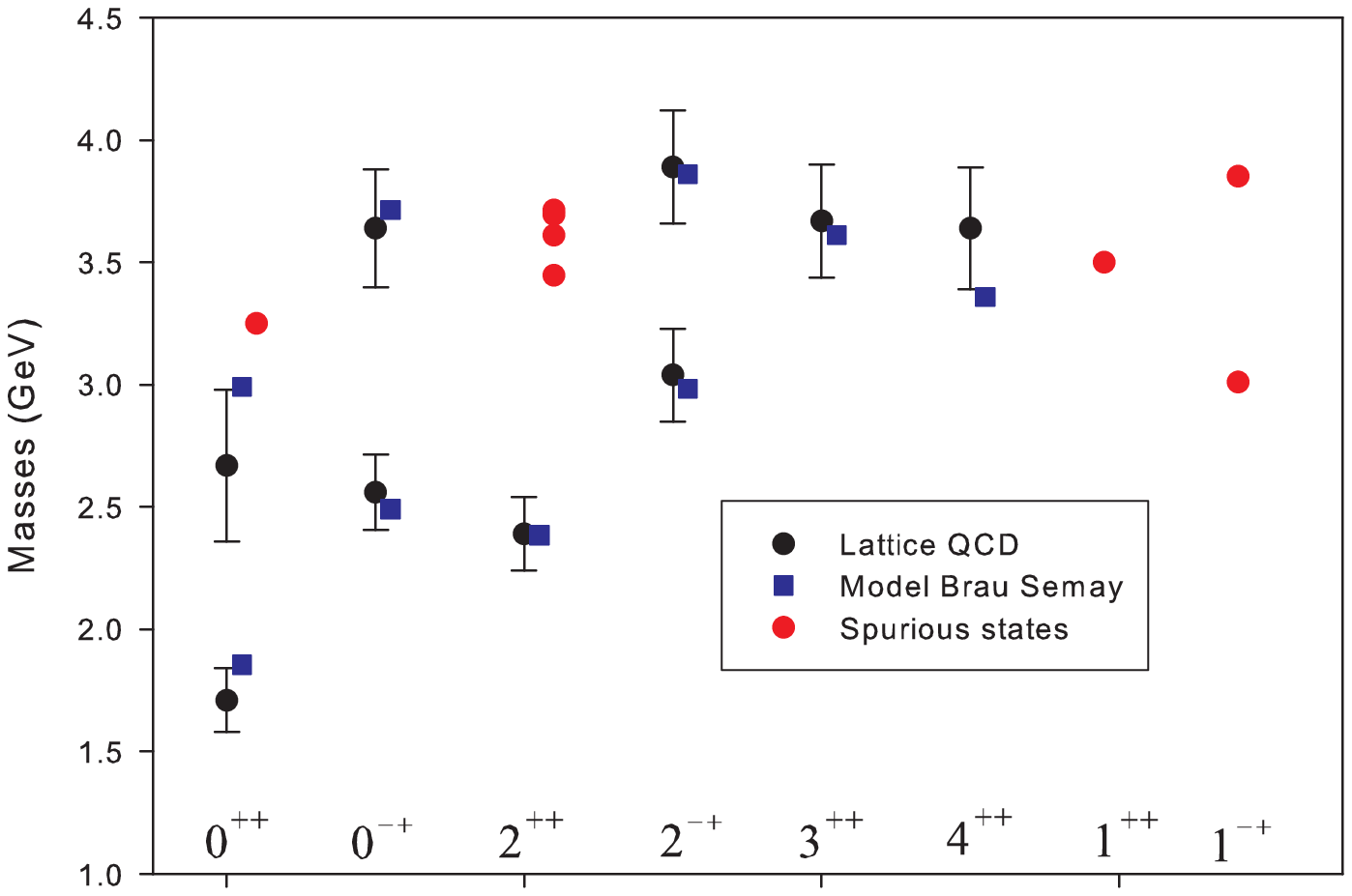}}
\caption{Left, glueball spectrum of ref.~\cite{Cornwall:1982zn}; right, comparison between lattice results~\cite{Morningstar:1999rf}~(circles) and two-gluon glueballs from ref.~\cite{Brau:2004xw}.\label{fig:2gluons}}
\end{figure}

The screened potential led to a spectrum with relatively low
glueball masses. The scalar and tensor glueballs had masses around
$\sim 1.3$ and $\sim 1.6$~GeV, respectively, see Fig~\ref{fig:2gluons}. At that time, there were no lattice study of the Yang-Mills spectrum and no conclusion concerning the relevance of the number of gluon degrees of freedom of were drawn.

After this pioneer work, various models were constructed on the same ground~\cite{Kaidalov:1999yd,Brau:2004xw} with the possibility to compare the predictions of the models with an accurate glueball spectrum obtained in quarkless QCD lattice  simulations~\cite{Morningstar:1999rf}.

All ingredients needed to reproduce the low-lying pure gauge spectrum of lattice QCD with two constituent gluons owning three degrees of freedom were identified~\cite{Brau:2004xw}. We display in Fig.~\ref{fig:2gluons} (right), the spectrum obtained in ref.~\cite{Brau:2004xw}. The Hamiltonian has a relativistic kinetic energy, $E_k = 2\sqrt{\bm p^2}$ and spin-dependent potentials coming from the OGE. We observed vector and tensor glueballs in the spectrum not observed in any lattice study but the others states match onto the lattice data.

At this stage, we can not conclude on the relevance of this model. A comparison with an equivalent model but with only two polarizations for the gluons is requested (cf. section~\ref{sec:2gluonhelicity}). The lattice studies~\cite{Morningstar:1999rf} offer us the possibility to extend the model for negative $C-$parity. In view of the results of ref.~\cite{Brau:2004xw}, we hope to reproduce the glueball spectrum with three gluons with an extension of this model. We will investigate this generalization in the next subsection.

\subsection{Three-gluon Glueballs}

A complete investigation of the glueball spectrum in constituent
models has to include three-gluon glueballs. Indeed, in this
approach negative $C$-parity glueballs involve at least three
constituents. There are two color wave functions, totally symmetric
or totally antisymmetric, coupling three adjoint representations
into a singlet. They do not mix and we are only interested in the
symmetric one, $d_{abc}A^a_\mu A^b_\nu A^c_\rho$, namely the $C=-$
states. The construction of the basis is a straightforward generalization of the two-body case. The only subtlety is the symmetry. Indeed, for the coupling of three spin$-1$ representations, we have
\begin{equation}\label{}
    \bm1\otimes\bm1\otimes\bm1 = \bm0_A\oplus\bm1_S\oplus\bm2_{MS}\oplus\bm3_S.
\end{equation}
The subscripts $A,S,MS$ stand for anti-symmetric, symmetric and mixed symmetry spin functions respectively. The total wave function is fully symmetric, the low-lying states ($L=0$) are then $1^{--},3^{--}$, and $0^{-+}$. We are are not interested in the latter since two gluons can also bound on positive charge conjugation. A state with the quantum numbers $2^{--}$ is only possible with a spacial function owing a mixed symmetry which will rise the mass of the state.

\begin{figure}[htb]
\includegraphics[width=0.5\linewidth]{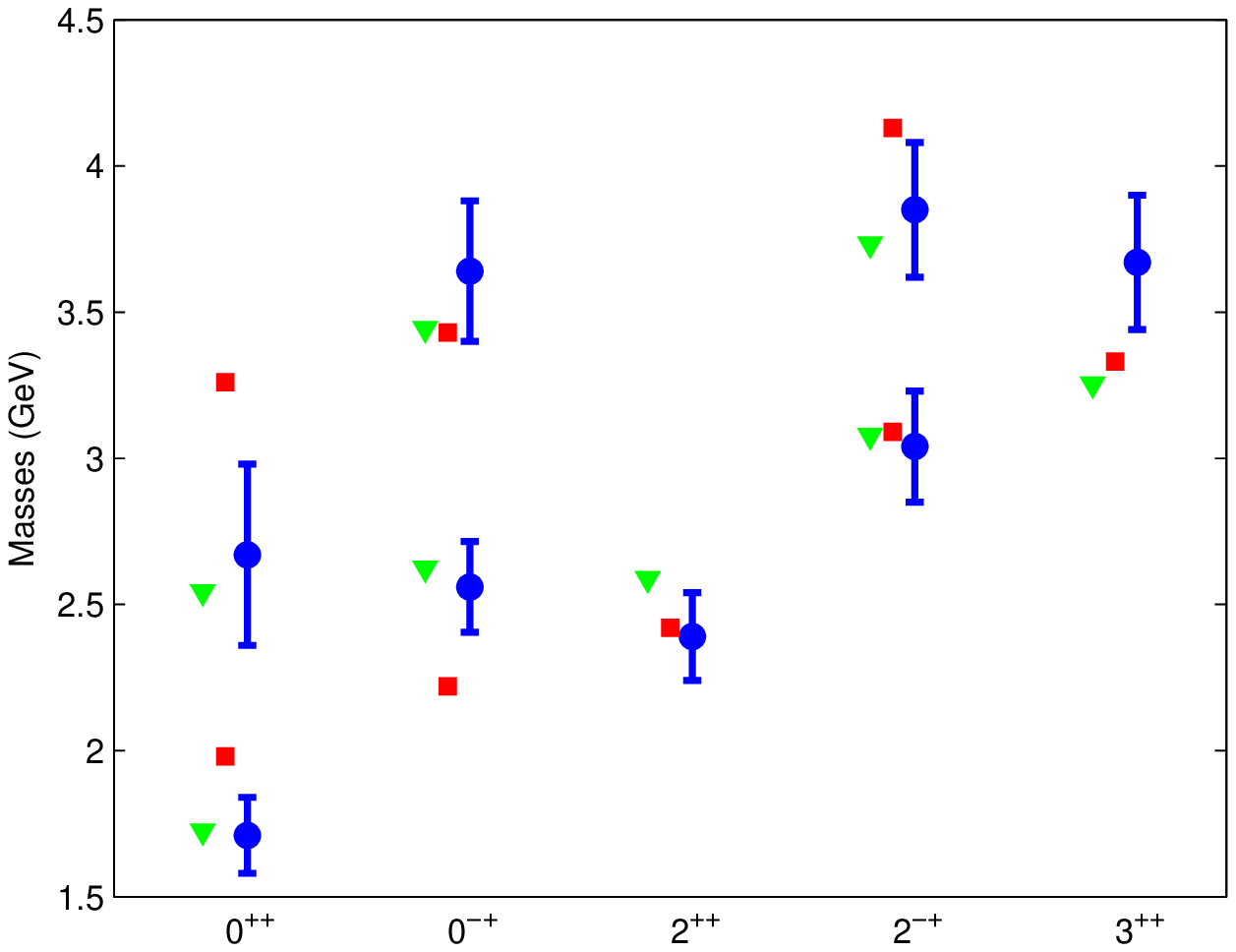}
\includegraphics[width=0.5\linewidth]{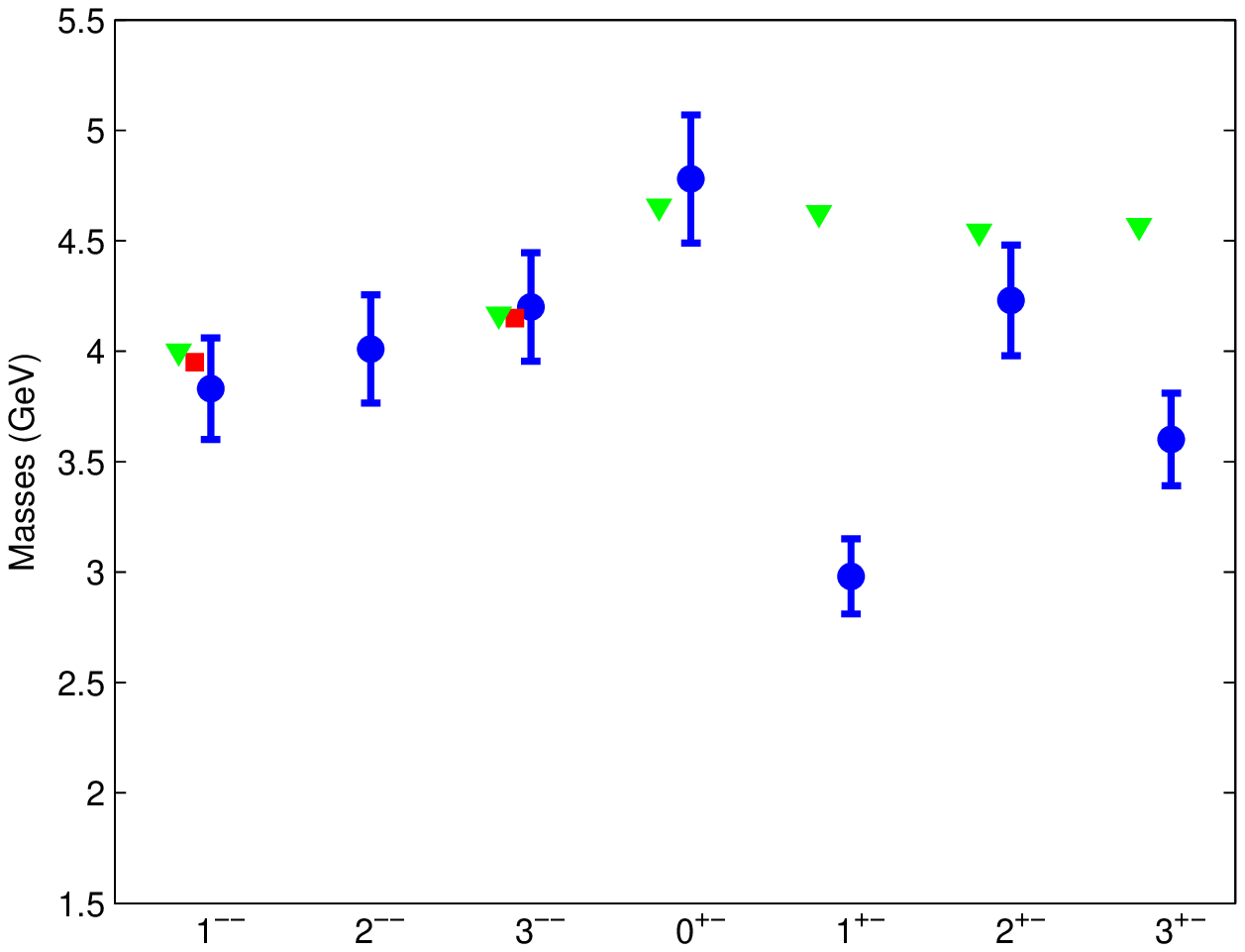}
\caption{Comparison between lattice
results~\cite{Morningstar:1999rf}~(circles) and two-gluon glueballs with two degrees of freedom (left) from ref.~\cite{Szczepaniak:1995cw}~(squares) and
ref.~\cite{Mathieu:2008bf}~(triangles) and three-gluon glueballs with three degrees of freedom (right) from ref.~\cite{LlanesEstrada:2005jf}~(squares) and
ref.~\cite{Mathieu:2008pb}~(triangles).\label{fig:3gluons}}
\end{figure}

The low-lying three-gluon states were studied in~\cite{Hou:1982dy,LlanesEstrada:2005jf,Mathieu:2006bp,Kaidalov:2005kz}. Only the first work~\cite{Hou:1982dy} utilized a non-relativistic Hamiltonian but all references considered spin-dependent potentiel coming from the OGE. Without the OGE potential, states with the same orbital angular momentum would be degenerate. The low-lying $1^{--}$ and $3^{--}$ lie in the same mass range, as expected from the symmetry argument, and are found in agreement with the lattice results, see Fig.~\ref{fig:3gluons} (right). From the lattice studies~\cite{Morningstar:1999rf}, we observed the $2^{--}$ lying between the $1^{--}$ and the $3^{--}$, in clear contradiction with the symmetry argument. Some authors~\cite{LlanesEstrada:2005jf,Kaidalov:2005kz} reproduced the tensor $2^{--}$ exactly where the lattice studies observed it. However, they made a mistake since they considered the same symmetry function for the three states $(1,2,3)^{--}$ ! When using the correct symmetry function, the $2^{--}$ appears to be higher the other states~\cite{Mathieu:2006bp}.

We pointed out a first discrepancy, the $2^{--}$, between lattice QCD and constituent models for glueballs when the gluon has three degrees of freedom. We now turn our attention to orbitally excited states. A positive parity requires an odd angular momentum. The coupling of the two internal angular momenta forming the main component $L=1$ has more than one symmetry. Hence, the four states $(0,1,2,3)^{+-}$ are degenerate without spin splitting~\cite{Mathieu:2008pb}. Even with the addition of spin-spin and spin-orbit interactions, they remain in the same mass range. There is no way to accommodate the lattice spectrum in which they are 1.8 GeV between the scalar and the vector, see Fig.~\ref{fig:3gluons} (right).

In contradiction with other authors, which missed the correct symmetry argument, we conclude in ref.~\cite{Mathieu:2008pb} that a constituent model for negative $C-$parity glueballs based on gluons with three degrees of freedom is not able to reproduce the lattice spectrum. However, for positive $C-$parity glueballs, constituent models in which the gluon has a third longitudinal component reproduce the lattice spectrum. But we notice the appearance of spurious states, vector and tensor, not observed in any lattice studies. In order to solve this apparent contradiction, we will now explore models with only two degrees of freedom for the gluon.

\section{Two Degrees of Freedom}
\subsection{Two-gluon Glueballs}
\label{sec:2gluonhelicity}
When constructing a basis with two spin-1 particles, the decomposition $\bm J=\bm L+\bm S$ is
employed. One then realizes that the basis contains too many states, for instance with
$J=1$~\cite{Brau:2004xw}. Those states are not obtained by any lattice study. A way out is to recall that the coupling of two transverse particles forbids the existence of vector states. This fact is known as Yang's theorem~\cite{Yang:1950rg} (a vector meson does not decay into two photons). One may then assume that, despite the fact that the gluon gains a mass (induced by confinement), it remains transverse with only two spin projections.

The formalism to treat two-body relativistic scattering developed by Jacob and Wick~\cite{Jacob:1959at} allows also the description of representations with only
transverse gluons. I sketch its main features and apply this formalism to the study of the
two-gluon glueball.

The Jacob and Wick formalism is based on states, $\left|J,M;\lambda_1,\lambda_2\right\rangle$,
which are eigenstates of $\bm J^2$ and $J_z$ and where $\lambda_1$ and $\lambda_2$ represent
the allowed spin projections. In the case under consideration, the projections can only be
maximal, {\it i.e.} $\pm s$ for a particle with helicity-$s$. The angular part of these
states are related with the conventional basis states by means of Clebsch-Gordan coefficients:
\begin{equation}\label{eq:decomp}
    \left|J,M;\lambda_1,\lambda_2\right\rangle=\sum_{L,S}
    \left[\frac{2L+1}{2J+1}\right]^{1/2}
    \left\langle LS0\Lambda\right|J\Lambda\left.
    \right\rangle\left\langle s_1s_2\lambda_1(-\lambda_2)
    \right|S\Lambda\left.
    \right\rangle\, \left|^{2S+1}L_J\right\rangle,
\end{equation}
with $\Lambda=\lambda_1-\lambda_2$. The radial part (depending on $J$) is determined
variationally with the Hamiltonian. The relation~\eqref{eq:decomp} is essential to compute matrix elements.

The states are not eigenstates of parity and does not have a good behavior under the exchange of the two particles. For a two-gluon state, $s_1=s_2=1$, it holds
\begin{eqnarray}\label{pdef}
    {\cal P}\left|J,M;\lambda_1,\lambda_2\right\rangle&=&(-1)^{J}
    \left|J,M;-\lambda_1,-\lambda_2\right\rangle\\
    {\mathrm P}_{12}\left|J,M;\lambda_1,\lambda_2\right\rangle&=&
(-1)^{J}\left|J,M;\lambda_2,\lambda_1\right\rangle.
\end{eqnarray}

A system of two gluons has to be totaly symmetric. Imposing this constraint as well as defined parity lead to selection rules on the spin. It can be checked that one can only obtain the
following states~\cite{Mathieu:2008bf}
\begin{equation}\label{ggstate}
        \left|S_+;(2k)^+\right\rangle,\quad \left|S_-;(2k)^-\right\rangle,\quad
        \left|D_+; (2k+2)^+\right\rangle,\quad \left|D_-; (2k+3)^+\right\rangle,
        \quad k\in\mathbb{N}.
\end{equation}
The $S$- and $D$-labels stand for helicity-singlet ($\lambda_1=\lambda_2$) and -doublet ($\lambda_1=-\lambda_2$) respectively. We recognized in Eq.~\eqref{ggstate} the four families
predicted by Yang for two photons~\cite{Yang:1950rg}.

It is readily observed that only the $\left|S_\pm;(2k)^+\right\rangle$ states can lead to $J=0$, while the $\left|D_\pm\right\rangle$ states always have $J\geq 2$ (since $J>|\lambda_1-\lambda_2|$). Obviously, a consequence of Yang's theorem is that no $J=1$ states are present. Only the $\left|D_-\right\rangle$ states can generate an odd-$J$, but $J$ is
at least 3 in this case.

Lattice QCD confirms the absence of the $1^{-+}$ and $1^{++}$
states, at least below $4$ GeV. It is worth mentioning that glueball
states with even-$J$ and positive parity can be built either from
the helicity-singlet or from the helicity-doublet states. The
important result is that the gluons remain transverse and therefore
the helicity formalism exactly reproduces the $J^{PC}$ content for
glueballs which is observed in lattice QCD, without the extra states
which are usually present in potential models with gluon owing three degrees of freedom.

The helicity formalism was applied for the first time by
Barnes~\cite{Barnes:1981ac}. It has several advantages not shared by
the more conventional non-relativistic $LS$-basis. It avoids
spurious states forbidden by the coupling of two transverse gluons
but also reproduces the lattice QCD hierarchy.

Within this approach, a given $J^{PC}$ state can be expressed as a linear combination of
$(L,S)$ states thanks to Eq.~\eqref{eq:decomp}. The complete expressions for these
decompositions can be found in Mathieu {\it et al}~\cite{Mathieu:2008bf}. We give here the
angular dependence, thanks to Eq.~\eqref{eq:decomp} of the ground states of Eq.~\eqref{ggstate}:
\begin{subequations}\label{eq:examples_states}
\begin{eqnarray}
\left|S_+;(0)^+\right\rangle&=&\sqrt{\frac{2}{3}}\left|^1
S_{0}\right\rangle +\sqrt{\frac{1}{3}}\left|^5 D_{0}\right\rangle,\\
    \left|S_-;(0)^-\right\rangle&=&\left|^3 P_{0}\right\rangle,\\
\left|D_+;(2)^+\right\rangle&=&\sqrt{\frac{2}{5}}\left|^5S_{2}\right\rangle
+\sqrt{\frac{4}{7}}\left|^5 D_{2}\right\rangle+\sqrt{\frac{1}{7}}\left|^5 G_{2}\right\rangle,\\
        \left|D_-;(3)^+\right\rangle&=&\sqrt{\frac{5}{7}}\left|^5D_{3}\right\rangle
    +\sqrt{\frac{2}{7}}\left|^5 G_{3}\right\rangle.
\end{eqnarray}
\end{subequations}
These decompositions are essential for computing the matrix elements
of non-relativistic operators (spin-spin, spin-orbit and tensor). Let us
note that the matrix elements of these operators are equal for
$\left|S_+;(2k)^+\right\rangle$ and
$\left|S_-;(2k)^-\right\rangle$~\cite{Mathieu:2008bf}. Therefore, the OGE potential would only furnish an overall contribution.

Even though in this approach the singlet states
$J^{P}=(2k^{+},2k^{-})$ are degenerate, with a  Cornell-type (linear
+ Coulomb) potential, a nonrelativistic kinetic energy, which
incorporates an {\it ad hoc} gluon mass $m$, and using the helicity
formalism, Barnes was able to reproduce the qualitative feature of
the pure gauge sector finding $M(0^{\pm+})=4.36 \, m$. The higher mass
ratios were not in perfect agreement with modern lattice results,
implying the need for modifications.

This improvement was carried out in a work based on the Coulomb
gauge Hamiltonian where a relativistic kinetic energy was
used~\cite{Szczepaniak:1995cw}. In this model, gluons are linked by
an adjoint string. The adjoint string tension $\sigma_A=(9/4)\sigma$
is expressed in terms of the well-known fundamental string tension
for mesons $\sigma$ through the Casimir scaling hypothesis. Using typical values for
the parameters, $\sigma=0.18$~GeV$^2$ for the fundamental string
tension (extracted from mesons Regge trajectory) and $\alpha_S=0.4$
for the strong coupling,  this model encodes the essential features
of glueballs.

The spectrum of the Coulomb gauge Hamiltonian was in good agreement
with lattice QCD. Moreover, the singlet $2^{-+}$ and $2^{++}$ are
degenerate as in the Barnes'~model, a characteristic of the helicity
formalism. The authors found a little difference between the scalar and
pseudoscalar glueball masses. This splitting, about 250~MeV was
nevertheless not as strong as in lattice QCD (850~MeV).

Recently,  this problem was revisited keeping the basic ingredients
needed for obtaining an acceptable pure gauge spectrum compatible
with lattice results, {\it i.e.} semi-relativistic energy and the
helicity formalism for two transverse gluons~\cite{Mathieu:2008bf}. A
simple Cornell potential was used but an instanton induced force was
added and with it the splitting between the scalar and pseudoscalar
glueballs was reproduced. There are arguments favoring an attractive
(repulsive) interaction induced by instantons in the scalar
(pseudoscalar) channel of glueballs~\cite{Schafer:1994fd}.

The two-gluon spectrum is displayed in Fig.~\ref{fig:3gluons} (left). We note that, comparing to the previous models (with the three degrees of freedom for the gluons), we have in the case at hand a perfect agreement with the lattice spectrum without the need of spin-depend potential. The states are naturally non degenerate thanks to the decompositions~\eqref{eq:examples_states}. Moreover, we have exactly the same quantum numbers as observed in lattice QCD without extra states. Finally, a instanton contribution is needed in the (pseudo)scalar sector as requested by the low energy theorems of the corresponding correlators~\cite{Mathieu:2008me,Schafer:1994fd,Forkel:2003mk}.

\subsection{Three-gluon Glueballs}
The conclusion of the previous sections requires to be supported by a three-gluon analysis with only two degrees of freedom. As far as I know, a proper inclusion of the helicity formalism in a three-gluon system was not already proposed. But developments in that direction was published~\cite{wick3,gie,gross}. The missing point is the generalization of the decomposition~\eqref{eq:decomp}. Such a formula would allow us to compute matrix elements between three-body wave functions. The intrinsic difficulty inherent to a many-body system is the number of angular momenta implying the multiplicity of recoupling schemes. Fortunately, the recoupling coefficients for the change of basis from $|(ij)k$ to $|(jk)i$ are known~\cite{wick3,gie}. In principle, applying the decomposition~\eqref{eq:decomp} between the cluster $(ij)$ and the particle $k$ and using the recoupling coefficients to fully symmetrize the wave function, would give the desired three-gluon wave functions. I expected that a proper inclusion of the three-body helicity formalism would reproduce the lattice spectrum for negative charge conjugation sector.

\section{Understanding the Lattice Spectrum}
\subsection{Glueball Correlators}
In the previous sections, I showed that the lattice pure gauge spectrum are better reproduce when considering only two gluonic degrees of freedom in a constituent models. I will now explain how this fact can be understood by some group theoretical arguments~\cite{Boulanger:2008aj}.

Glueball masses are extracted from the corresponding correlators. For numerical purposes, a discrete version of the correlators is implemented on a lattice but our classification remains valid. The arguments developed here are also valid for other techniques (sum rules, AdS/QCD, etc.)~\cite{Mathieu:2008me} used to find the properties of the gluonic correlators.

The gluon field strength $F_{\mu\nu}$ is antisymmetric and has six components. But with the equations of motion, $p^{\mu}F_{\mu\nu}=0$ and $p^{\mu}\widetilde F_{\mu\nu}=0$, only two components survive on-shell. To see this, it is instructive to use the light-cone coordinates, $p^\pm = (p^0\pm p^3)/\sqrt{2}$, in which $p^\mu=(E,0,0,E)$ has only one non-zero component, $p^+=E\sqrt{2}$\footnote{with the light-cone metric, $g_{-+}=g_{+-}=g_{ii}=1$, one has $p^{-}=p_+$.}. It is then straightforward to see that the only two independent components are $F_{-i}=p_{-}A_i$ with $i\in\{1,2\}$. Correlators are then build out of gluons with two degrees of freedom.

The low-lying glueball correlators are composed of two gluon field strengths. Projecting out the color singlet, we obtain the following decomposition in irreducible tensors
\begin{equation}\label{decompo2F}
    F_{\mu\nu}^a\otimes F_{\alpha\beta}^b\stackrel{\delta_{ab}}{\longrightarrow} F_{\mu\nu}^aF^{\mu\nu}_a\oplus F_{\mu\nu}^a\widetilde{F}^{\mu\nu}_a \oplus \left(F_{\mu\alpha}^aF_{\nu}^{a\alpha}-\frac{1}{4}g_{\mu\nu}F_{\alpha\beta}^aF^{\alpha\beta}_a \right) \oplus
    \left(F_{\mu\alpha}^a\widetilde F_{\nu}^{a\alpha}-\frac{1}{4}g_{\mu\nu} F_{\alpha\beta}^a \widetilde F^{\alpha\beta}_a\right).
\end{equation}
The ground states of two gluons is then composed of a scalar, a pseudoscalar, a tensor and a axial tensor. They correspond exactly to the states observed in lattice QCD and obtained with the helicity formalism. The excited states of two gluons are obtained in a similar way but starting with $F_{\mu\nu}^a D_{[\lambda_1}\cdots D_{\lambda_n]}F_{\alpha\beta}^a$~\cite{Jaffe:1985qp}.

The decomposition~\eqref{decompo2F} can be recast in a more convenient form by the use of young diagrams~\cite{Boulanger:2008aj}. We denote by a box $\yng(1)$ a gluon with two degrees of freedom. Upper indices represent color indexes. We have to symmetrize the product of two boxes since the two particles are identical and we get
\begin{equation}\label{y1}
\left( \yng(1)^{a} \otimes \yng(1)^{b} \right) =  \bullet^{(ab)}\oplus \yng(2)^{(ab)} \oplus \yng(1)^{[ab]}
\end{equation}
The vector state is not a color singlet and disappear when we contract the tensors with the singlet color function. We can repeat the same operation for the low-lying three-gluon states:
\begin{equation}\label{y3}
\left( \yng(1)^{a} \otimes \yng(1)^b \otimes \yng(1)^{c} \right) = \bullet^{[abc]} \oplus \yng(1)^{(abc)}\oplus \yng(1)^{\tiny\young(ac,b)}\oplus \yng(2)^{\tiny\young(ac,b)} \oplus \yng(3)^{(abc)}.
\end{equation}
Only the totaly symmetric $(abc)$ and antisymmetric $[abc]$ color functions give rive to color singlets. They are respectively $C=-$ and $C=+$. From the decomposition in irreducible tensors~\eqref{y3}, we see that the only low-lying three-gluons with a negative charge conjugation are the vector and the spin 3 tensor. Those quantum numbers should be the ones we would get by an application of the three-body helicity formalism. But there are exactly the low-lying glueballs with negative $C-$parity observed in lattice QCD.

\subsection{A Four Gluon State}
The lattice studies tell us that beside the low-lying spin 1 and spin 3, there are two spin 2 and one scalar glueballs with $C=-$. We could understand the spin 2 as orbital excitation of the lower spin 1. However, we claimed that the scalar $0^{+-}$ is actually a four-gluon states~\cite{Boulanger:2008aj}. An orbital excitation can not give rise to a scalar particle. Moreover, there are two scalars in the decomposition of four gluons
\begin{equation}\label{y4}
( \yng(1)^{a} \otimes \yng(1)^{b} \otimes \yng(1)^{c} \otimes \yng(1)^{d}\, )
= \bullet^{(abcd)}\oplus \bullet^{\tiny\young(ab,cd)} \oplus \cdots,
\end{equation}
where the ellipses denotes higher spin tensors. The configuration $\tiny \young(ab,cd)$ corresponds to an negative $C-$parity in which the gluon are in the color function $[[\bm 8,\bm 8]^{\overline{\bm{10}}},[\bm 8,\bm 8]^{\bm{10}}]^{\bm{1}}$~\cite{Boulanger:2008aj}. In this particular configuration, the Coulomb interaction vanished between each pair of gluons. The interaction is only given by the confinement.

In the ref.~\cite{Boulanger:2008aj}, we computed the mass of the $0^{+-}$. Since we are not able to implement the helicity formalism for more than two particles, we assumed a decomposition in two clusters of two gluons. This approximation led us to a (dimensionless) mass
\begin{equation}\label{}
    r_0M_{0^{+-}} = 11.61,
\end{equation}
in agreement with the recent lattice result $r_0M_{0^{+-}}=11.66\pm0.19$~\cite{Morningstar:1999rf}.

\subsection{Gluelump Spectrum}
A gluelump is defined as a bound state of the gluonic field and a static (scalar) octet color source. The gluelump spectrum was computed on a lattice in ref.~\cite{gl2}. The low-lying gluelumps can be interpreted in term of a single gluon bound with the octet source~\cite{Guo:2007sm}. When the gluon has only degrees of freedom, the hierarchy observed in lattice is exactly reproduced, {\it i.e} $1^{+-}$, $1^{--}$, $2^{--}$, $2^{+-}$, $3^{+-}$. However, if we allow the gluon to have the longitudinal component, the lowest gluelump state is a $0^{--}$~\cite{Boulanger:2008aj}. This argument supports the idea developed in the manuscript that the gluon has only two degrees of freedom although it gains an effective mass.

\section{Conclusion}
I reviewed two different constituent models of glueballs with the aim to identify the correct number of gluonic degrees of freedom in quarless QCD. Models with three gluonic degrees of freedom are apparently in agreement with the low-lying (positive charge conjugation) lattice spectrum. However, the existence of extra states in the basis, the need of spin-dependent forces and the complete forgetting of instanton-induced interactions in the scalar channels are strong arguments not in favor of this approach. Indeed, the extension to negative charge conjugation (three-gluon glueballs) is in clear disagreement with the lattice spectrum. The symmetry arguments invoked in the demonstration revealed the misconstruction of the basis.

The problems encountered in the first model disappear if we use gluons with only two degrees of freedom. The construction of the basis is a little bit more involved but the exact hierarchy without spurious states is recovered. The interaction potential can be taken at its lowest form (a Cornell shape without OGE) and instanton contributions are included as requested by other techniques. Within this model clear understanding of the lattice spectrum is gained. It remains nevertheless to generalize to three-gluon states. But I am convince that a proper inclusion of the helicity formalism in a Hamiltonian approach of glueballs would agree with the $C=-$ lattice spectrum.

Massive representations have more degrees of freedom that massless ones. In the case at hand, a longitudinal component would be naively added for a massive interpretation of a gluon. However, in view of the results outlined above, the gluon behaves intrinsically a massless particle although it obeys a massive dispersion relation.

In quantum field theory, bound states are described by interpolating currents. Since the currents are build out of gluon fields (or field strengths) with only two degrees of freedom, it is then natural to include only two degrees of freedom for the constituent gluons in effective models. I always compared to lattice results due to the lack of clear experimental candidates. I concluded that only two gluonic degrees of freedom are needed to reproduce numerical simulations of correlators but it remains only valid for quarkless QCD. The inclusion of quarks induces mixing with mesons states with all the problems it implies~\cite{Mathieu:2008me,Crede:2008vw}. But, with or without quark, if we still believe in the description of bound states in term of interpolating currents, we have to consider two gluonic degrees of freedom in effective approaches although the gluon might gain a mass.

\section*{Acknowledgements}
I thank J. Papavassiliou who suggested me to review this topic. I also enjoyed several discussions with the participants of the workshop in particular with P.~Bicudo, J. Cornwall, D.~Dudal, O.~Oliviera, V.~Sauli, N.~Vandersickel.

\end{document}